\documentclass{iau_FM}
\usepackage{graphicx}

\title%% give here short title %%
{Graphene and Carbon Nanotubes in Space}

\author[Chen, Xiao, Li \& Zhong] %% give here short author list %%
{Xiuhui~Chen$^1$, Zichun Xiao$^2$, Aigen Li$^3$,  \and Jianxin~Zhong$^1$}

\pubyear{2018}
\jname{Astronomy in Focus, Volume 2} 
\affiliation{$^1$Xiangtan University, China;
$^2$Ravenscroft School, Raleigh, NC 27615, USA;
$^3$University of  Missouri, USA;
email: {\tt lia@missouri.edu}}
\editors{Maria Teresa V.T. Lago, ed.}
\begin{document}

\maketitle
As the fourth most abundant element 
in the universe, carbon plays an important 
role in the physical and chemical evolution 
of the interstellar medium (ISM).
Due to its unique property to form three 
different types of chemical bonds through 
sp$^{1}$, sp$^{2}$, and sp$^{3}$ hybridizations, 
carbon can be stabilized in various allotropes, 
including amorphous carbon, graphite, 
diamond, polycyclic aromatic hydrocarbon (PAH),
fullerenes, graphene, and carbon nanotubes (CNTs).

Many allotropes of carbon are known to be present 
in the ISM (Henning \& Salama 1998).
Presolar graphite grains and nanodiamonds
have been identified in primitive meteorites based 
on their isotopically anomalous composition
(see Nittler 2018). While hydrogenated
amorphous carbon grains reveal their 
presence in the diffuse ISM through the ubiquitous 
3.4\,$\mu$m aliphatic C--H absorption feature
(Pendleton \& Allamandola 2002),
the aromatic C--H and C--C emission features
at 3.3, 6.2, 7.7, 8.6 and 11.3\,$\mu$m
infer the widespread presence of PAHs 
in a wide variety of interstellar regions
(see Hudgins \& Allamandola 2005).
The detections of interstellar C$_{60}$ 
and C$_{70}$ (Cami et al.\ 2010, 
Sellgren et al.\ 2010)
and their cations
(Bern\'e et al.\ 2013,
Strelnikov et al.\ 2015)
have also been reported 
based on their characteristic
infrared (IR) emission features.

\begin{figure*}
\centerline{
\includegraphics[scale=0.85,clip]{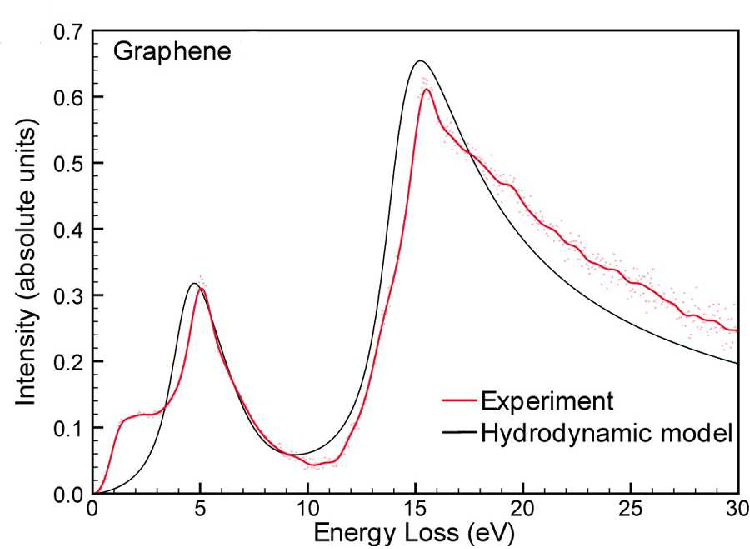}\hspace{0.4cm}
\includegraphics[scale=0.65,clip]{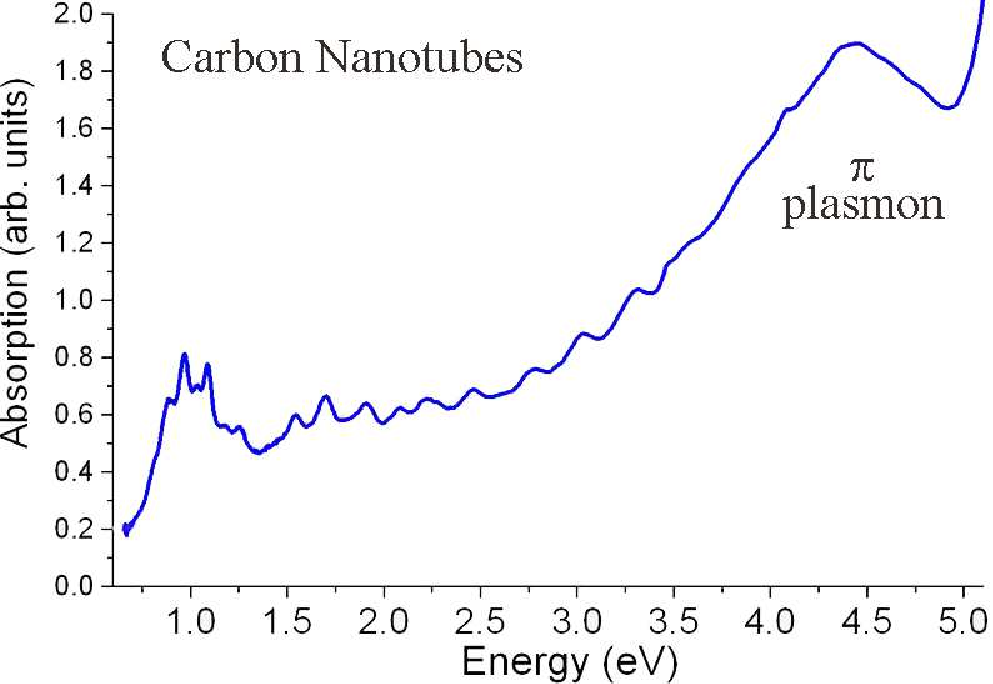}
}
\caption{\footnotesize
       \label{fig:GraCNT}
       {\it Left} (a): The energy-loss function (ELF) 
       of graphene, obtained experimentally 
       and calculated using a two-dimensional 
       hydrodynamic model by Nelson et al.\ (2014),
       is dominated by two peaks at $\sim$\,4.5 
       and $\sim$\,15\,eV, known as $\pi$ 
       and $\pi$+$\sigma$ plasmons, respectively. 
       {\it Right} (b): The optical absorption spectrum 
       from dispersed single-wall CNTs 
       measured by Kataura et al.\ (1999) 
       exhibits a number of electronic transitons, 
       with the $\pi$ plasmon also peaking 
       at $\sim$\,4.5\,eV being the most prominent.
       The nondetection in the Galactic interstellar 
       extinction curve of the $\sim$\,4.5\,eV 
       absorption peak of graphene and CNTs
       would allow one to place an upper limit
       on the amounts of graphene and CNTs
       in the ISM.
       }
\end{figure*}

Graphene was first experimentally 
synthesized in 2004 
by A.K.~Geim and K.S.~Novoselov
for which they received 
the 2010 Nobel Prize in physics.
More recently, 
Garc{\'{\i}}a-Hern{\'a}ndez et al.\ (2011, 2012)
reported for the first time the presence of
unusual IR emission features at $\sim$\,6.6, 9.8, 
and 20\,$\mu$m in several planetary nebulae,
both in the Milky Way and the Magellanic Clouds,
which are coincident with the strongest transitions 
of planar C$_{24}$, a piece of graphene.
In principle, graphene could be present 
in the ISM as it could be formed from
the photochemical processing of PAHs,
which are abundant in the ISM,
through a complete loss of 
their H atoms (e.g., see Bern\'e \& Tielens 2012).
On the other hand,
as illustrated in Figure~\ref{fig:GraCNT}a,
both quantum-chemical computations 
and laboratory experiments have shown 
that the exciton-dominated 
$\pi$--$\pi^{\ast}$
electronic transitions in graphene
cause a strong absorption band 
near 2755\,{\rm \AA} or 4.5\,eV
(Yang et al.\ 2009, Nelson et al.\ 2010)
which is not seen in the ISM.
This allows us to place an upper limit
of $\sim$\,5\,ppm of C/H on the abundance
of graphene in the diffuse ISM
(see Chen et al.\ 2017).
Moreover, the nondetection of
the 6.6, 9.8, and 20\,$\mu$m emission features
of graphene C$_{24}$ in the observed IR 
emission spectra of the diffuse ISM
is also consistent with an upper limit of
$\sim$\,5\,ppm of C/H 
in the C$_{24}$ graphene sheet
(see Chen et al.\ 2017).

%%% Figure 1 %%%

%%%%%%%%%% CNTs %%%%%%%%%%%%

CNTs can be envisioned as a layer of 
graphene sheet rolled up into a cylinder.
%Since their discovery in early 1990s 
%by Iijima (1991), CNTs have attracted 
%considerable interest worldwide 
%because of their unusual properties 
%(e.g., optical activity, circular dichroism)
%and great potentials for technological applications. 
%
%The structure of a CNT is completely specified 
%by the chiral vector which is given in terms of 
%a pair of integers ($n,m$). 
%CNTs can be classified according to
%whether the tube is {\it achiral}
%--- ``zigzag'' with ($n,0$) 
%and ``armchair'' with ($n,n$), 
%or {\it chiral} with ($n$, $m$) 
%but $n$\,$\neq$\,$m$.
%Chiral nanotubes can exhibit very large 
%(i.e., long) one-dimensional (1D) unit cells 
%compared to achiral tubes of the same diameter.
%
%Electrically, CNTs can be either metallic 
%or semiconducting, depending on their 
%geometry, i.e., on ($m,n$). 
%Armchair CNTs and zigzag CNTs 
%with $m$\,=\,3$q$
%(where $q$ is an integer)
%are always metallic.
%
%
%Single-walled CNTs (SWNTs) 
They
are novel 1D materials made of 
an sp$^2$-bonded wall one atom thick. 
Strong confinement ($\sim$\,1\,nm) 
of charge carriers 
%in SWNTs 
results in 
their unique optical properties being
dominated by strongly bound excitons 
%(Coulomb-bound electron-hole pairs) 
as revealed by the sharp optical absorption features.
%
%
%and photoluminescence peaks 
%in the near-IR region.
%(M. J. O’Connell, S. M. Bachilo, C. B. Huffman,  et al.\ 2002,
%Science, 297, 593)
%
It would be interesting to explore
whether CNTs could be responsible for
some of the mysterious diffuse interstellar bands
(e.g., see Zhou et al.\ 2006).
%and the unidentified ``extended red emission''
%(ERE) which is an interstellar photoluminescence
%phenomenon (Witt \& Vijh 2004). 

As illustrated in Figure~\ref{fig:GraCNT}b,
CNTs also exhibit a broad and intense 
absorption feature at $\sim$\,4.5\,eV that is 
typically attributed to a $\pi$-plasmon
excitation (e.g., see Kataura et al.\ 1999),
although the exact position of this feature
appears to vary with nanotube diameter
(see Rance et al.\ 2010).
Similar to graphene, 
the absence of the $\sim$\,4.5\,eV 
absorption in the ISM
would allow us to place an upper limit
on the abundance of interstellar CNTs.
%(see Chen et al.\ 2018).

Like graphene, 
CNTs would emit in the IR
through their C--C vibrational modes.
CNTs are actually more IR-active 
than graphene due to 
their cylindrical boundary condition.
Many vibrational modes 
in the $\sim$\,680--1730\,cm$^{-1}$ 
range have been experimentally
detected for CNTs (e.g., see Kim et al.\ 2005).
%
%
%
%Compared to a graphene sheet
%for which the crystal symmetry forbids 
%a $q$\,=\,0 IR active optical phonon,
%one expects more IR active vibrational 
%modes for CNTs since this rule is broken
%by their cylindrical boundary conditions.
%Indeed, the IR-active vibrational modes 
%of CNTs at $\sim$\,682, 806, 854, 1033, 
%1095, 1262, 1369, 1443, 1564, 
%and 1730\,cm$^{-1}$ have been detected 
%by optical transmission through thin films 
%of bundled nanotubes (Kim et al.\ 2005).
{\it JWST}'s unique high sensitivity and
high resolution IR capabilities will 
open up an IR window unexplored by 
{\it Spitzer} and unmatched by {\it ISO} observations 
%due to the sensitivity difference
and thus will allow us to explore 
the possible presence of graphene 
and CNTs in the ISM in greater detail.

\vspace{1mm}
\noindent
{\it Acknowledgements}:~We are supported in part by
NSF AST-1816411
and an AAS international travel grant. 

\end{document}